\documentclass[conference]{IEEEtran}
\usepackage[bookmarks=false]{hyperref}
\usepackage{amsthm}
\usepackage{quantum}
\usepackage{tikz}
\usetikzlibrary{}
\usepgfmodule{oo}
\allowdisplaybreaks[4]

\begin{document}

\pgfdeclarelayer{background}
\pgfdeclarelayer{firstbackground}
\pgfdeclarelayer{secondbackground}
\pgfsetlayers{secondbackground,firstbackground,background,main}

\title{An improved rate region for the classical-quantum broadcast channel}

\author{
    \IEEEauthorblockN{Christoph Hirche\IEEEauthorrefmark{1} and Ciara Morgan\IEEEauthorrefmark{2}}
    \IEEEauthorblockA{Institut f\"{u}r Theoretische Physik\\ Leibniz Universit\"{a}t Hannover\\ Appelstra\ss e 2, D-30167 Hannover, Germany\\
Email: \IEEEauthorrefmark{1}christoph.hirche@itp.uni-hannover.de \IEEEauthorrefmark{2}ciara.morgan@itp.uni-hannover.de}
}

\maketitle

\begin{abstract}
We present a new achievable rate region for the two-user binary-input classical-quantum broadcast channel. The  result is a generalization of the classical Marton-Gelfand-Pinsker region and is provably larger than the best previously known rate region for classical-quantum broadcast channels. The proof of achievability is based on the recently introduced polar coding scheme and its generalization to quantum network information theory. 
\end{abstract}


\section{Introduction}

One of the fundamental tasks in quantum information theory is to determine the maximum possible rate at which information can be sent reliably from one party to another over a noisy communication channel. Indeed to show the \emph{achievability} of a certain rate, one needs to prove the existence of a code, that is, an encoding and decoding scheme, that achieves this rate with vanishing error in the limit of many channel uses. 

Network information theory centers on the study and analysis of communication rates in the multi-user setting, generalizing the single sender and receiver case. 
The broadcast channel is one the most fundamental channels in this field and, in the two-user case, it models the simultaneous communication between a single sender and two receivers. In the classical setting, there exist several schemes to prove that certain rate regions are achievable for the broadcast channel. Two of these schemes, which are of particular interest to us, are known as superposition coding \cite{B73, C72} and binning \cite{M79}, also called multicoding. For the binning scheme, independent messages are sent simultaneously to both receivers and can be decoded by the respective receiver, according to an argument based on joint typicality. For the superposition scheme we exploit the fact that certain inputs are decodable by both receivers. Marton \cite{M79}  combined both techniques in order to send private messages over the two-user broadcast channel, achieving the region now known as Marton's region. Later this result was extended by Gelfand and Pinsker \cite{GP80} where a common message can be sent to both receivers, resulting in the so-called Marton-Gelfand-Pinsker region with common messages. Interestingly, it can be shown, in the classical setting, that even when we set the common rate in the Marton-Gelfand-Pinsker region to zero the resulting rate region is, in some cases, larger than Marton's region \cite{GK11}. 

Broadcast channels have also been generalized to the setting of classical-quantum communication \cite{YHD11, SW12, RSW14}. 
To date the best known rate region for the classical-quantum channel has been established in \cite{SW12, RSW14} and it is a generalization of Marton's region in the classical case.

Arikan \cite{A09} recently introduced the now celebrated polar coding scheme for classical channels. Indeed Arikan showed that these codes can achieve the symmetric capacity of any classical single-sender single-receiver channel in the limit of many channel uses and, remarkably, that this can be done with a complexity $O(N\log N)$ for encoding and decoding, where $N$ is the number of channels used for communication. 
Moreover, polar codes make use of the effect of channel polarization, where a recursive construction is used to divide the instances of a channel into a fraction that can be used for reliable communication and a fraction that is nearly useless. The crucial feature is that the fraction of \emph{good} channels is approximately equal to the symmetric capacity of the channel. 

Polar codes have attracted great deal of attention and were generalized to many additional communication settings, such as the task of source coding \cite{A10, A12} and universal coding for compound channels \cite{HU13, MELK132}.

Polar codes have also been generalized to the setting of sending classical information over quantum channels \cite{RDR12,WG13,WG12}, in addition to sending quantum information \cite{WR12, WG13QDeg, SRDR13}. For the task of sending classical information, in addition to asymmetric channels \cite{H14}, the quantum setting has also been generalized to certain multi-user channels, namely the multiple access and interference channels and compound multiple access channel ~\cite{HMW14, H14}. 
Recently polar codes have also been applied to the classical broadcast channel \cite{GAG13, MHSU14}. 
In \cite{MHSU14} the authors show how polar codes can be used for the broadcast channel to achieve the Marton-Gelfand-Pinsker region with and without common messages. 

In this work we will show that the approach of \cite{MHSU14} can be used to achieve the Marton-Gelfand-Pinsker region with and without common messages for classical-quantum broadcast channels, giving rise to the largest known rate region for the classical-quantum broadcast channel. 

The remainder of this work is organized as follows. In Section \ref{notation} we will state the necessary preliminaries and in Section \ref{cqBroadcast} we show how to achieve the Marton-Gelfand-Pinsker region for the broadcast channel using polar codes, before we conclude in Section \ref{Conclusion}. 

\section{Notation and definitions}\label{notation}

We begin by introducing certain notation which will be used throughout the article, before defining necessary entropic quantities and measures. 

In the remaining work $u_1^N\equiv u^N$ will denote a row vector $(u_1, \dots, u_N)$ and correspondingly $u_i^j$ will denote, for $1\leq i, j\leq N$, a subvector $(u_i, \dots, u_j)$.
Note that if $j<i$ then $u_i^j$ is empty. Similarly, for a vector $u_1^N$ and
a set $A \subset \{ 1,\dots, N\}$, we write $u_A$ to denote the subvector $%
(u_i : i\in A)$.

A discrete classical-quantum channel $W$ takes realizations $x \in \mathcal{X%
}$ of a random variable $X$ to a quantum state, denoted $\rho_x^B$, on a
finite-dimensional Hilbert space $\mathcal{H}^B$, 
\begin{equation}
W : x \rightarrow \rho_{x}^{B},
\end{equation}
where each quantum state $\rho_x$ is described by a positive semi-definite
operator with unit trace. We will take the input alphabet $\mathcal{X} =
\{0,1\}$ unless otherwise stated, and the tensor product $W^{\otimes N}$ of $%
N$ channels is denoted by $W^N$.

To characterize the behavior of symmetric classical-quantum channels, we will make use of the
symmetric Holevo capacity, defined as follows: 
\begin{equation}
I(W) \equiv I(X;B)_\rho,
\end{equation}
where the quantum mutual information with respect to a classical-quantum
state $\rho^{XB}$ is given by  
\begin{equation}
I(X;B) \equiv H(X)_\rho + H(B)_\rho - H(XB)_\rho,
\end{equation}
with $\rho^{XB} = \frac{1}{2} | 0\rangle\!\langle 0 | \otimes \rho_0^B + 
\frac{1}{2} | 1\rangle\!\langle 1 | \otimes \rho_1^B.$
In the above, the von Neumann entropy $H(\rho)$ is defined as 
$H(\rho) \equiv -\tr\{\rho \log_2 \rho\}.$
We will also make use of the conditional entropy defined as 
$H(X| B)_\rho = H(X)_\rho - H(XB)_\rho$
and the quantum conditional mutual information defined
for a tripartite state $\rho^{XYB}$ as 
$I(X;B| Y)_\rho \equiv H(XY)_\rho + H(YB)_\rho - H(Y)_\rho - H(XYB)_\rho.$

We characterize the reliability of a channel $W$ as the fidelity between the
output states 
\begin{equation}
F(W) \equiv F(\rho_0, \rho_1),
\end{equation}
with $F(\rho_0, \rho_1) \equiv \normTr{\sqrt{\rho_0}\sqrt{\rho_1}}^2$
and $\normTr{A} \equiv \tr{\sqrt{A^\dagger A}}.$
Note that in the case of two commuting density matrices, $\rho = \sum_i p_i \ketbra{}{i}{i}$ and $\sigma = \sum_i q_i \ketbra{}{i}{i}$ the fidelity can be written as $F(\rho, \sigma) = \left(\sum_i \sqrt{p_i q_i}\right)^2.$
Note that, the Holevo capacity and the fidelity can be seen as quantum generalizations
of the mutual information and the Bhattacharya parameter from the classical
setting, respectively (see, e.g., \cite{A09}).

We will also use the quantity 
\begin{equation}\label{Z}
Z(X| B)_{\rho} \equiv 2\sqrt{p_0p_1}\, F(\rho_0, \rho_1),
\end{equation}
introduced in \cite{SRDR13}, for a classical-quantum state $\rho$ which can, again, be seen as quantum generalization of the Bhattacharya parameter, for a classical variable, now with quantum side information. 

We will now define the two-user classical-quantum broadcast channel. 
The broadcast channel can
be modeled mathematically as the triple $\left( \mathcal{X},W,\mathcal{H}^{B_{1}}\otimes \mathcal{H}^{B_{2}}\right),$ with 
\begin{equation}
W : x \rightarrow  \rho_{x}^{B_1B_2}.
\end{equation}
The information processing task for the two-user classical-quantum broadcast channel is described as follows. The sender
would like to communicate messages to both
receivers. These messages are independent, but can also contain some common part for both receivers. The model is such that the first receiver only has access to the output system $B_1$ and therefore receives $\rho^{B_1}_x = \tr_{B_2} \rho^{B_1B_2}_x$, similarly, the second receiver has $\rho^{B_2}_x = \tr_{B_1} \rho^{B_1B_2}_x$. The sender chooses a message $m_k$ for each receiver from a message set $\mathcal{M}_k = \{1, \cdots, 2^{nR_k} \}$, and encodes her messages with the resulting the codeword $x^n(m_1,m_2) \in \mathcal{X}^n$.
The receivers' corresponding decoding POVMs are denoted by $\{\Lambda _{m_{1}}\}$ and $\{\Gamma _{m_{2}}\}$.
The code is said to be a $(n,R_{1},R_{2},\epsilon )$-code, if the average
probability of error is bounded as follows 
\begin{equation}
\bar{p}_{e}=\frac{1}{|\mathcal{M}_{1}||\mathcal{M}_{2}|}%
\sum_{m_{1},m_{2}}p_{e}(m_{1},m_{2})\leq \epsilon ,
\end{equation}%
where the probability of error $p_{e}(m_1, m_2)$ for a pair of messages $(m_{1},m_{2})$
is given by 
\begin{equation}
p_{e}(m_{1},m_{2})=\mathrm{Tr}\left\{\left( I-\Lambda _{m_{1}}\otimes \Gamma
_{m_{2}}\right) \rho _{x^n(m_1,m_2)}^{B^n_{1}B^n_{2}}\right\},
\end{equation}
with $\rho _{x^n(m_1,m_2)}^{B^n_{1}B^n_{2}}$ the state resulting
when the sender transmits the codeword $x^n(m_1,m_2)$ 
through $n$ instances of the channel. A rate pair $(R_1, R_2)$ is said to be \emph{achievable} for the two-user
classical-quantum broadcast channel described above if there exists an $(n, R_1,R_2, \epsilon)$-code $\forall \epsilon >0$ and large enough $n$.

\subsection{Polar codes for asymmetric channels}\label{asy}

In this section we will review polar codes for achieving the capacity of asymmetric channels, for more details we refer to \cite{H14}. 
Essentially polar codes are described by a linear transformation given by $x^N=u^NG_N$, where $u^N$ is the input sequence and 
\begin{equation}
G_N = B_NF^{\otimes n}
\end{equation}
with 
\begin{equation}
F \equiv \left[ 
\begin{matrix}
1 & 0 \\ 
1 & 1
\end{matrix}
\right],
\end{equation}
and $B_N$ is a permutation matrix known as a ``bit reversal'' operation \cite{A09}. 
This is called channel combining and transforms $N$ single copies of a channel $W$ to a channel $W_N$. 

In the second step, called channel splitting, $W_N$ is used to define $W^{(i)}_N$ as follows:
\begin{equation}
W^{(i)}_N : u_i \rightarrow \rho_{(i),u_i}^{U_1^{i-1}B^N},
\end{equation}
where 
\begin{equation}
\rho^{U^{i-1}_1 B^N}_{(i),u_i} = \sum_{u_1^{i-1}} \frac{1}{2^{i-1}} %
\ketbra{}{u_1^{i-1}}{u_1^{i-1}} \otimes \sum_{u_{i+1}^N} \frac{1}{2^{N-i}}
\rho^{B^N}_{u^N}.
\end{equation}
This is equivalent to a decoder which estimates, by the $i$-th measurement, the bit $u_i$, with the following
assumptions: the entire output is available to the decoder, the previous
bits $u_1^{i-1}$ are correctly decoded and the distribution over the bits $u^N_{i+1}$ is uniform.
The assumptions that all previous bits are correctly decoded is called ``genie-aided'' and can be ensured by a limited amount of classical communication prior to the information transmission. 
The decoder described above is thus a ``genie-aided'' successive cancellation decoder. 

These two steps give rise to the effect of channel polarization, which ensures that the fraction of channels $%
W_N^{(i)}$ which have the property $I(W_N^{(i)}) \in (1-\delta, 1]$ goes to
the symmetric Holevo information $I(W)$ and the fraction with $I(W_N^{(i)})
\in [0, 1-\delta)$ goes to $1-I(W)$ for any $\delta \in (0,1)$, as $N$ goes to
infinity through powers of two. This is one of the main insights of the work by Arikan \cite{A09} and the generalization in \cite{WG13} to the classical-quantum setting (see \cite{WG13} for a more detailed statement). 
To achieve the symmetric capacity we can now simply send information bits over the channels $I(W_N^{(i)}) \in (1-\delta, 1]$ and send prearranged ``frozen'' bits over the remaining channels. 

It turns out that, the above approach essentially works for asymmetric channels as well, the crucial point to see this, is that the polar coding transform $G_N$ is its own inverse for binary inputs. 
We consider the reverse protocol of classical lossless compression. Hence we can use a uniformly distributed input sequence and transform it to a distribution suitable to achieve the asymmetric capacity of the channel. Due to a polarization effect we can use a fraction of size $H(X)$ for the following channel coding. Note that in the special case of a symmetric channel the uniformity of the required input distribution simply gives a fraction $H(X)=1$. 
It is shown in \cite{H14} that this approach achieves the asymmetric capacity 
\begin{equation}
C(W) = \max_{p(x)} I(X;B) . 
\end{equation}
This is a generalization of a result in \cite{HY13} to the classical-quantum setting. 

\subsection{Alignment of polarized sets}\label{alignment}
From the definition of polar codes, it is clear that the set of channels which can be used for information transmission depend on the particular communication channel to be used. 
Polarizing, in a scenario with multiple possible channels, such as the case of compound channels, where one must code at rates which are achievable for all channels in a particular known a set of channels, will hence, lead to the situation where some synthesized channels are good for one channel but not for another, and vice versa. This problem can be solved by the technique of \emph{alignment} \cite{HU13}, which is described in detail for the classical-quantum compound channel in \cite{H14}. The main idea is to combine the channels which are good in one case with the channels good in the other by additional CNOT gates. Doing this recursively, we can halve the number of incompatible indices in every step. With the number of channels uses approaching infinity, we can minimize the number of incompatible indices.

\section{Broadcast channel} \label{cqBroadcast}

\subsection{Marton-Gelfand-Pinsker region for private messages}\label{MGPpm}

In this section we will show how to achieve the Marton-Gelfand-Pinsker region, initially without the use of common messages for classical-quantum broadcast channels using polar codes. 
Indeed, we will use the technique of alignment as explained in Section \ref{alignment} to achieve the rate region
\begin{align}
\begin{split}\label{marton}
R_1 &\leq I(V,V_1;B_1), \\
R_2 &\leq I(V,V_2;B_2) \\
R_1 + R_2 &\leq I(V,V_1;B_1) + I(V_2;B_2|V) - I(V_1;V_2|V), \\
R_1 + R_2 &\leq I(V,V_2;B_2) + I(V_1;B_1|V) - I(V_1;V_2|V),
\end{split}
\end{align}
for the classical-quantum two-user broadcast channel described by a classical input $X = \varphi(V, V_1, V_2)$ and a quantum output $\rho_x^{B_1B_2}$. 

Let $V, V_1, V_2$ be auxiliary binary random variables with $(V, V_1, V_2) \sim p_V p_{V_2|V} p_{V_1|V_2V}$. Now, let $X = \varphi(V, V_1, V_2)$ be a deterministic function. 
Without loss of generality we consider a broadcast channel such that $I(V; B_1) \leq I(V;B_2)$. 
With $G_n$ the usual polar coding transformation, set
\begin{align}
U_{(0)}^{n} = V^{n}G_n, \\
U_{(1)}^{n} = V_1^{n}G_n, \\
U_{(2)}^{n} = V_2^{n}G_n.
\end{align}
The variable $U_{(1)}^{n}$ carries the message of the first user, while $U_{(0)}^{n}$ and $U_{(2)}^{n}$ carry the message of the second user. 
To exploit the technique of superposition, we take $U_0^n$ to be decodable by both receivers, while $V$ carries information only for the second receiver. The variables $V_1, V_2$ correspond to the binning scheme and can only be decoded by one receiver, respectively.  

To handle these additional auxiliary variables we use the polarization technique used to achieve the asymmetric capacity of a channel introduced in Section \ref{asy}. Hence we introduce sets to determine the polarization of the probability distribution for the input and the channel independently. 
Define for $l\in\{ 1,2\}$, the following sets, with interpretations provided below,
\begin{align*}
\HH_{V} &= \{ i\in [n] : Z(U_{(0),i}\mid U_{(0)}^{n-1} ) \geq \delta_n \}, \\
\LL_{V}  &= \{ i\in [n] : Z(U_{(0),i}\mid U_{(0)}^{n-1} ) \leq \delta_n \},  \\
\HH_{V|B_l} &= \{ i\in [n] : Z(U_{(0),i}\mid U_{(0)}^{n-1} B_{(l)}^{n} ) \geq \delta_n \}, \\
\LL_{V|B_l}  &= \{ i\in [n] : Z(U_{(0),i}\mid U_{(0)}^{n-1} B_{(l)}^{n} ) \leq \delta_n \}, \\
\HH_{V_l|V} &= \{ i\in [n] : Z(U_{(l),i}\mid U_{(l)}^{n-1} U_{(0)}^{n} ) \geq \delta_n \}, \\
\LL_{V_l|V}  &=  \{ i\in [n] : Z(U_{(l),i}\mid U_{(l)}^{n-1} U_{(0)}^{n}  ) \leq \delta_n \}, \\
\HH_{V_l|V,B_l} &= \{ i\in [n] : Z(U_{(l),i}\mid U_{(l)}^{n-1} U_{(0)}^{n} B_{(l)}^{n} ) \geq \delta_n \}, \\
\LL_{V_l|V,B_l}  &=  \{ i\in [n] : Z(U_{(l),i}\mid U_{(l)}^{n-1} U_{(0)}^{n} B_{(l)}^{n}  ) \leq \delta_n \}, \\
\HH_{V_1|V,V_2} &= \{ i\in [n] : Z(U_{(1),i}\mid U_{(1)}^{n-1} U_{(0)}^{n} U_{(2)}^{n} ) \geq \delta_n \}, \\
\LL_{V_1|V,V_2}  &=  \{ i\in [n] : Z(U_{(1),i}\mid U_{(1)}^{n-1} U_{(0)}^{n} U_{(2)}^{n}  ) \leq \delta_n \},
\end{align*}
which due to the polarization effect satisfy
\begin{align*}
\lim_{n\rightarrow\infty} \frac{1}{n} |\HH_{V}| &= H(V), \\
\lim_{n\rightarrow\infty} \frac{1}{n} |\LL_{V}| &= 1 - H(V), \\
\lim_{n\rightarrow\infty} \frac{1}{n} |\HH_{V|B_l}| &= H(V|B_l), \\
\lim_{n\rightarrow\infty} \frac{1}{n} |\LL_{V|B_l}| &= 1 - H(V|B_l), \\
\lim_{n\rightarrow\infty} \frac{1}{n} |\HH_{V_l|V}| &= H(V_l|V), \\
\lim_{n\rightarrow\infty} \frac{1}{n} |\LL_{V_l|V}| &= 1 - H(V_l|V), \\
\lim_{n\rightarrow\infty} \frac{1}{n} |\HH_{V_l|V,B_l}| &= H(V_l|V,B_l), \\
\lim_{n\rightarrow\infty} \frac{1}{n} |\LL_{V_l|V,B_l}| &= 1 - H(V_l|V,B_l), \\
\lim_{n\rightarrow\infty} \frac{1}{n} |\HH_{V_1|V,V_2}| &= H(V_1|V,V_2), \\
\lim_{n\rightarrow\infty} \frac{1}{n} |\LL_{V_1|V,V_2}| &= 1 - H(V_1|V,V_2).
\end{align*}
Note that the polarization of the classical quantities follows from polar codes for classical source coding \cite{A10}, while the polarization of the quantities with quantum side information is shown in \cite{SRDR13}. 

Intuitively $\HH_{V}$ and $\LL_{V}$ describe the polarization of the random variable $V$ and correspond to whether or not the $i$th bit is nearly completely deterministic given the previous bits, respectively. 
Similarly $\HH_{V|B_l}$ and $\LL_{V|B_l}$ determine whether the $l$th receiver can decode bits knowing the previous inputs and all outputs. 
$\HH_{V_l|V}$, $\LL_{V_l|V}$, $\HH_{V_l|V,B_l}$, $\LL_{V_l|V,B_l}$ have the same interpretation, with the additional side information from first decoding $V$. 
$\HH_{V_1|V,V_2}$ and $\LL_{V_1|V,V_2}$ handle the indices decoded by the first user, with assumed knowledge of $V$ and $V_2$, while in our case that user does not have access to the latter. 

Recall that $U_{(0)}^{n}$ can be decoded by both users but only contains information for the second one. Define $\II^{(2)}_{sup} = \HH_V \cap \LL_{V|B_2}$ to contain positions decodable for the second user and $\II_v^{(1)} = \HH_V \cap \LL_{V|B_1}$ those decodable for the first user. 
Again, $U_{(2)}^{n}$ can only be decoded by the second user and also only contains information for this receiver. Indeed by $\II^{(2)}_{bin} = \HH_{V_2|V} \cap \LL_{V|B_2}$ we denote the set of indices which can be decoded by the second receiver. 
$U_{(1)}^{n}$ only needs to be decoded by the first user and also only contains information for that user. The set $\II^{(1)} = \HH_{V_1|V} \cap \LL_{V|B_1}$ denotes the indices which the first receiver can decode reliably. 
We also have to take into account that the first user cannot decode $U_{(2)}^{n}$ therefore the indices in the set $\FF^{(1)} = \LL_{V_1|V,V_2} \cap \HH_{V_1|V} \cap \HH_{V_1|V,B_1}$ are critical. 

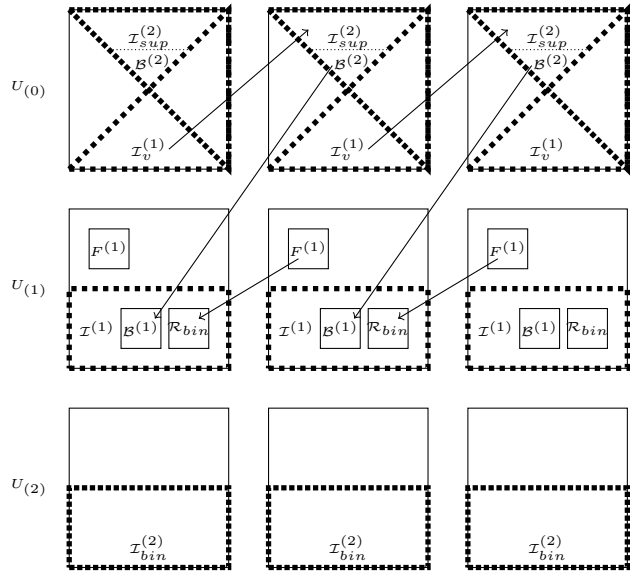
\begin{figure*}[t]\label{broadcast}
\centering
\begin{tikzpicture}[scale=0.53]
\tikzstyle{every node}=[font=\tiny]
	\node[] at (0+2,10+0.5) {$\II_v^{(1)}$};
	\node[] at (0+2,10+3.3) {$\II^{(2)}_{sup}$};
	\node[] at (5+2,10+0.5) {$\II_v^{(1)}$};
	\node[] at (5+2,10+3.3) {$\II^{(2)}_{sup}$};
	\node[] at (10+2,10+0.5) {$\II_v^{(1)}$};
	\node[] at (10+2,10+3.3) {$\II^{(2)}_{sup}$};
	\draw[line width=2pt, dotted] (0,10) -- (0+4,10+4) -- (0+4,10) -- cycle; 
	\draw[line width=2pt, dotted] (5,10) -- (5+4,10+4) -- (5+4,10) -- cycle; 
	\draw[line width=2pt, dotted] (10,10) -- (10+4,10+4) -- (10+4,10) -- cycle; 
	\draw[line width=2pt, densely dotted] (0,14) -- (0+4,10+4) -- (0+4,10) -- cycle; 
	\draw[line width=2pt, densely dotted] (5,14) -- (5+4,10+4) -- (5+4,10) -- cycle; 
	\draw[line width=2pt, densely dotted] (10,14) -- (10+4,10+4) -- (10+4,10) -- cycle; 
	\node at (0+2,0+0.5) {$\II^{(2)}_{bin}$};
	\node at (5+2,0+0.5) {$\II^{(2)}_{bin}$};
	\node at (10+2,0+0.5) {$\II^{(2)}_{bin}$};
	\draw[line width=2pt, densely dotted] (0,0) -- (0+4,0+0) -- (0+4,2) -- (0,2) -- cycle; 
	\draw[line width=2pt, densely dotted] (5,0) -- (5+4,0+0) -- (5+4,2) -- (5,2) -- cycle; 
	\draw[line width=2pt, densely dotted] (10,0) -- (10+4,0+0) -- (10+4,2) -- (10,2) -- cycle; 
	\draw[line width=2pt, dotted] (0,5) -- (0+4,5+0) -- (0+4,5+2) -- (0,5+2) -- cycle; 
	\draw[line width=2pt, dotted] (5,5) -- (5+4,5+0) -- (5+4,5+2) -- (5,5+2) -- cycle; 
	\draw[line width=2pt, dotted] (10,5) -- (10+4,5+0) -- (10+4,5+2) -- (10,5+2) -- cycle; 
	\myssstamp.box(0,0,2);
	\myssstamp.box(5,0,2);
	\myssstamp.box(10,0,2);
	\myssstamp.box(0,5,1);
	\myssstamp.box(5,5,1);
	\myssstamp.box(10,5,1);
	\myssstamp.box(0,10,0);
	\myssstamp.box(5,10,0);
	\myssstamp.box(10,10,0);
           \draw[->,solid] (2.5,10.5)  -- (6,13.5); 
           \draw[->,solid] (7.5,10.5)  -- (11,13.5); 
	\draw[] (0.5,7.5) -- (1.5,7.5) -- (1.5,8.5) -- (0.5,8.5) -- cycle;
	\draw[] (5.5,7.5) -- (6.5,7.5) -- (6.5,8.5) -- (5.5,8.5) -- cycle;
	\draw[] (10.5,7.5) -- (11.5,7.5) -- (11.5,8.5) -- (10.5,8.5) -- cycle;
	\draw[] (2.5,5.5) -- (3.5,5.5) -- (3.5,6.5) -- (2.5,6.5) -- cycle;
	\draw[] (7.5,5.5) -- (8.5,5.5) -- (8.5,6.5) -- (7.5,6.5) -- cycle;
	\draw[] (12.5,5.5) -- (13.5,5.5) -- (13.5,6.5) -- (12.5,6.5) -- cycle;
	\node at (1,8) {$\FF^{(1)}$};
	\node at (6,8) {$\FF^{(1)}$};
	\node at (11,8) {$\FF^{(1)}$};
	\node at (3,6) {$\RRR_{bin}$};
	\node at (8,6) {$\RRR_{bin}$};
	\node at (13,6) {$\RRR_{bin}$};
           \draw[->,solid] (11.6,12.6)  -- (7.15,6.25); 
           \draw[->,solid] (6.6,12.6)  -- (2.15,6.25); 
           \draw[->,solid] (10.75,7.75)  -- (8.25,6.25); 
           \draw[->,solid] (5.75,7.75)  -- (3.25,6.25); 
	\draw[] (1.3,5.5) -- (2.3,5.5) -- (2.3,6.5) -- (1.3,6.5) -- cycle;
	\draw[] (6.3,5.5) -- (7.3,5.5) -- (7.3,6.5) -- (6.3,6.5) -- cycle;
	\draw[] (11.3,5.5) -- (12.3,5.5) -- (12.3,6.5) -- (11.3,6.5) -- cycle;
	\node at (1.8,6) {$\BB^{(1)}$};
	\node at (6.8,6) {$\BB^{(1)}$};
	\node at (11.8,6) {$\BB^{(1)}$};
	\draw[densely dotted] (1.0,13) -- (3,13);
	\draw[densely dotted] (6.0,13) -- (8,13);
	\draw[densely dotted] (11.0,13) -- (13,13);
	\node at (2.1,12.7) {$\BB^{(2)}$};
	\node at (7.1,12.7) {$\BB^{(2)}$};
	\node at (12.1,12.7) {$\BB^{(2)}$};
	\node[] at (0.7,6) {$\II^{(1)}$};
	\node[] at (5.7,6) {$\II^{(1)}$};
	\node[] at (10.7,6) {$\II^{(1)}$};
	\node at (-1,12) {$U_{(0)}$};
	\node at (-1,7) {$U_{(1)}$};
	\node at (-1,2) {$U_{(2)}$};
\end{tikzpicture}
\caption[Coding for the broadcast channel.]{Coding for the broadcast channel. Indices in dotted subsets are considered to be good for a specific receiver denoted by the associated set. Arrows indicate the alignment process. For a colored version of this figure see \cite{H14}.}
\end{figure*}

We now use, in total, three different steps of alignment construction as described in Section \ref{alignment} and illustrated in Figure \ref{broadcast}. 
First we handle $U_{(0)}^{n}$. By definition these variables should be decoded by both receivers and contain information only for the second one. 
We simply use the alignment technique for classical-quantum channels to send the message assigned for the second receiver to both of them. By the assumption that $I(V; B_1) \leq I(V;B_2)$ we conclude that we can reliably send $I(V; B_1)$ of information to both users. We know that whenever $I(V; B_1)$ is not equal to $I(V; B_2)$ there will be unaligned indices remaining. Lets call this set $\BB^{(2)}$. 

In the second step we choose a subset $\BB^{(1)}$ of $\II^{(1)}$ such that $|\BB^{(1)}| = |\BB^{(2)}|$. We can then align these two subsets and therefore raise the number of indices from $U_{(0)}^{n}$, which both receivers can decode, to $I(V; B_2)$. 

In the third step we need to cope with the fact that the first user cannot decode the informations in $\FF^{(1)}$. 
Again choose a subset $\RRR_{bin}$ of $\II^{(1)}$ such that $|\RRR_{bin}| = |\FF^{(1)}|$. We use $\RRR_{bin}$ to repeat the information for the first user in $\FF^{(1)}$ of the following block. 

In order to get the correct order for the successive cancellation decoder, we will encode $U_{(0)}^{n}$ and $U_{(2)}^{n}$ forward, while $U_{(1)}^{n}$ will be decoded backwards. 
Moreover, the first receiver decodes $U_{(0)}^{n}$ and $U_{(1)}^{n}$ forwards, while the second receiver decodes $U_{(0)}^{n}$ and $U_{(2)}^{n}$ backwards. 
\pgfooclass{ssstamp}{ % This is the class ssstamp
    \method ssstamp() { % The constructor 
    }
 \method cnot(#1,#2,#3,#4) { % Causes the ssstamp to be shown at coordinate (#1,#2)
        %Draw the ssstamp:
	\draw (#1,#2) -- (#1-#3,#2) -- (#1-#3,#2-#4) -- (#1,#2-#4);
	\draw (#1-#3,#2) circle (0.1);
	\draw (#1-#3,#2+0.1) -- (#1-#3,#2) -- (#1-#3-0.25,#2);
	\draw[dotted] (#1-#3-0.25,#2) -- (#1-#3-0.75,#2);
   }
 \method box(#1,#2,#3) { % Causes the ssstamp to be shown at coordinate (#1,#2)
        %Draw the ssstamp:
	\draw (#1,#2) -- (#1+4,#2) -- (#1+4,#2+4) -- (#1,#2+4) -- cycle; 
   }
}
\pgfoonew \myssstamp=new ssstamp()

Now if we let the number of blocks approach infinity, we can calculate the rate for the first receiver as follows
\begin{align*}
R_1 &= \frac{1}{n} (|\II^{(1)}| - |\BB^{(1)}| - |\RRR_{bin}|) \\
&= I(V_1; B_1 \mid V) - I(V_1; V_2\mid V) - (I(V;B_2) - I(V;B_2)) \\
&= I(V,V_1; B_1) - I(V_1; V_2\mid V) - I(V;B_2). 
\end{align*}
We can also calculate the rate for the second receiver
\begin{align*}
R_2 &= \frac{1}{n}(|\II^{(2)}_{sup}| + |\II^{(2)}_{bin}|) \\
&= I(V; B_2) + I(V_2; B_2\mid V) \\
&= I(V,V_2;B_2). 
\end{align*}

Finally note that if we swap the role of the two receivers, the set $\BB^{(2)}$ will be empty due to the assumption that $I(V; B_1) \leq I(V;B_2)$, therefore we can achieve the rates
\begin{align}
R_1 &= I(V,V_1; B_1) \\
R_2 &= I(V_2; B_2\mid V) - I(V_1; V_2\mid V).
\end{align}

For the classical case it is known \cite{MHSU14} that these two rate pairs coincide with the Marton-Gelfand-Pinsker rate region. The proof can be directly translated to the setting of classical quantum communication and therefore our scheme achieves the rate region stated in Equation \ref{marton}.

\subsection{Marton-Gelfand-Pinsker region with common messages}

We can simply extend our coding scheme in Section \ref{MGPpm} to include the transmission of a common message for both receivers, by noting that the information sent via $U_{(0)}^{n}$ can be reliably decoded by both users. Therefore we can use these indices to send an amount of information equal to $\min\{ I(V;B_1), I(V;B_2)\}$ to both users. 
This leads to the rate region
\begin{align}
\begin{split}\label{macm}
R_0  &\leq \min\{ I(V;B_1), I(V;B_2)\}, \\
R_0 + R_1 &\leq I(V,V_1;B_1), \\
R_0 + R_2 &\leq I(V,V_2;B_2) \\
R_0 + R_1 + R_2 &\leq I(V,V_1;B_1) + I(V_2;B_2|V) - I(V_1;V_2|V), \\
R_0 + R_1 + R_2 &\leq I(V,V_2;B_2) + I(V_1;B_1|V) - I(V_1;V_2|V).
\end{split}
\end{align}

\section{Conclusion}\label{Conclusion}

We have shown that, using polar coding, it is possible to achieve a new rate region for classical-quantum broadcast channel, which coincides with the classical Marton-Gelfand-Pinsker region and is larger than the previously known rate regions for this channel. Hence this work gives an example where polar coding can be used to prove the achievability of new rate regions.  

\section*{Acknowledgments}
We would like to thank Mark M. Wilde for enjoyable and fruitful discussions.  
This work was supported by the EU grants SIQS and QFTCMPS and by the cluster of excellence EXC 201 Quantum Engineering and Space-Time Research.

\bibliographystyle{IEEEtran}
\bibliography{bib}

\begin{thebibliography}{10}
\providecommand{\url}[1]{#1}
\csname url@samestyle\endcsname
\providecommand{\newblock}{\relax}
\providecommand{\bibinfo}[2]{#2}
\providecommand{\BIBentrySTDinterwordspacing}{\spaceskip=0pt\relax}
\providecommand{\BIBentryALTinterwordstretchfactor}{4}
\providecommand{\BIBentryALTinterwordspacing}{\spaceskip=\fontdimen2\font plus
\BIBentryALTinterwordstretchfactor\fontdimen3\font minus
  \fontdimen4\font\relax}
\providecommand{\BIBforeignlanguage}[2]{{%
\expandafter\ifx\csname l@#1\endcsname\relax
\typeout{** WARNING: IEEEtran.bst: No hyphenation pattern has been}%
\typeout{** loaded for the language `#1'. Using the pattern for}%
\typeout{** the default language instead.}%
\else
\language=\csname l@#1\endcsname
\fi
#2}}
\providecommand{\BIBdecl}{\relax}
\BIBdecl

\bibitem{B73}
P.~P. Bergmans, ``Random coding theorem for broadcast channels with degraded
  components,'' \emph{IEEE Transactions on Information Theory}, vol.~19, no.~2,
  pp. 197--207, March 1973.

\bibitem{C72}
T.~Cover, ``Broadcast channel,'' \emph{IEEE Transactions on Information
  Theory}, vol.~18, no.~1, pp. 2--14, January 1972.

\bibitem{M79}
K.~Marton, ``A coding theorem for the discrete memoryless broadcast channel,''
  \emph{IEEE Transactions on Information Theory}, vol.~25, no.~3, pp. 306--311,
  May 1979.

\bibitem{GP80}
S.~I. Gelfand and M.~S. Pinsker, ``Capacity of a broadcast channel with one
  deterministic component,'' \emph{Problems of Information Transmission},
  vol.~16, no.~1, pp. 17--25, January 1980.

\bibitem{GK11}
A.~E. Gamal and Y.-H. Kim, \emph{Network Information Theory}.\hskip 1em plus
  0.5em minus 0.4em\relax Cambridge University Press, 2011.

\bibitem{YHD11}
J.~Yard, P.~Hayden, and I.~Devetak, ``Quantum broadcast channels,'' \emph{IEEE
  Transactions on Information Theory}, vol.~57, no.~10, pp. 7147--7162, October
  2011, arXiv:quant-ph/0603098.

\bibitem{SW12}
I.~Savov and M.~M. Wilde, ``Classical codes for quantum broadcast channels,''
  \emph{Proceedings of the 2012 IEEE International Symposium on Information
  Theory}, pp. 721--725, July 2012, arXiv:1111.3645.

\bibitem{RSW14}
J.~Radhakrishnan, P.~Sen, and N.~Warsi, ``{One-shot Marton inner bound for
  classical-quantum broadcast channel},'' October 2014, arXiv:1410.3248.

\bibitem{A09}
E.~Arikan, ``Channel polarization: A method for constructing capacity-achieving
  codes for symmetric binary-input memoryless channels,'' \emph{IEEE
  Transactions on Information Theory}, vol.~55, no.~7, pp. 3051--3073, July
  2009, arXiv:0807.3917.

\bibitem{A10}
------, ``{Source polarization},'' Jan. 2010, arXiv:1001.3087.

\bibitem{A12}
------, ``Polar coding for the {Slepian-Wolf} problem based on monotone chain
  rules,'' \emph{Proceedings of the 2012 IEEE International Symposium on
  Information Theory}, pp. 566--570, July 2012.

\bibitem{HU13}
S.~H. Hassani and R.~Urbanke, ``Universal polar codes,'' July 2013,
  arXiv:1307.7223.

\bibitem{MELK132}
H.~Mahdavifar, M.~El-Khamy, J.~Lee, and I.~Kang, ``Compound polar codes,''
  February 2013, arXiv:1302.0265.

\bibitem{RDR12}
J.~M. Renes, F.~Dupuis, and R.~Renner, ``Efficient polar coding of quantum
  information,'' \emph{Physical Review Letters}, vol. 109, no.~5, p. 050504,
  August 2012, arXiv:1109.3195.

\bibitem{WG13}
M.~M. Wilde and S.~Guha, ``Polar codes for classical-quantum channels,''
  \emph{IEEE Transactions on Information Theory}, vol.~59, no.~2, pp.
  1175--1187, February 2013, arXiv:1109.2591.

\bibitem{WG12}
S.~Guha and M.~M. Wilde, ``Polar coding to achieve the {Holevo} capacity of a
  pure-loss optical channel,'' in \emph{Proceedings of the 2012 IEEE
  International Symposium on Information Theory}, Cambridge, MA, USA, July
  546-550.

\bibitem{WR12}
J.~M. Renes and M.~M. Wilde, ``Polar codes for private and quantum
  communication over arbitrary channels,'' \emph{IEEE Transactions on
  Information Theory}, vol.~60, no.~6, pp. 3090--3103, June 2014,
  arXiv:1212.2537.

\bibitem{WG13QDeg}
M.~M. Wilde and S.~Guha, ``Polar codes for degradable quantum channels,''
  \emph{IEEE Transactions on Information Theory}, vol.~59, no.~7, pp.
  4718--4729, July 2013, arXiv:1109.5346.

\bibitem{SRDR13}
D.~Sutter, J.~M. Renes, F.~Dupuis, and R.~Renner, ``Efficient quantum polar
  codes requiring no preshared entanglement,'' \emph{Proceedings of the 2013
  IEEE International Symposium on Information Theory}, pp. 354--358, July 2013,
  arXiv:1307.1136.

\bibitem{H14}
C.~Hirche, ``{Polar codes in quantum information theory},'' 2014, {Master's
  thesis, Hannover, arXiv:1501.03737}.

\bibitem{HMW14}
C.~Hirche, C.~Morgan, and M.~M. Wilde, ``Polar codes in network quantum
  information theory,'' 2014, arXiv:1409.7246.

\bibitem{GAG13}
N.~Goela, E.~Abbe, and M.~Gastpar, ``Polar codes for broadcast channels,''
  \emph{Proceedings of the 2013 IEEE International Symposium on Information
  Theory}, pp. 1127--1131, 2013, arXiv:1301.6150.

\bibitem{MHSU14}
M.~Mondelli, S.~H. Hassani, I.~Sason, and R.~Urbanke, ``Achieving {Marton's}
  region for broadcast channels using polar codes,'' January 2014,
  arXiv:1401.6060.

\bibitem{HY13}
J.~Honda and H.~Yamamoto, ``Polar coding without alphabet extension for
  asymmetric models,'' \emph{IEEE Transactions on Information Theory}, vol.~59,
  no.~12, pp. 7829--7838, Dec. 2013.

\end{thebibliography}

\end{document}